\documentclass[a4paper,12pt]{article}
\usepackage{amssymb,amsthm,latexsym}

\title{About the Almeida-Thouless transition line in the 
Sherrington-Kirkpatrick mean field spin glass model}

\author{
Fabio Lucio Toninelli\footnote{\ 
e-mail: {\tt f.toninelli@sns.it}} \\
{\small {\itshape Scuola Normale Superiore, Piazza dei Cavalieri 7, 56126 Pisa,
Italy}}\\
{\small {\itshape and Istituto Nazionale di Fisica Nucleare, Sezione di 
Pisa}}
} 

\date{\today}

\begin{document}

\maketitle

In this short note, we consider the Sherrington-Kirkpatrick mean 
field spin glass model \cite{sk, sk2} and we prove that, in the 
thermodynamic limit $N\to\infty$, the quenched free energy per site is 
strictly greater than
the corresponding replica symmetric approximation \cite{sk}, for 
values of temperature and magnetic field below the Almeida-Thouless line 
\cite{at}.
This is a simple consequence of rigorous bounds, discovered  by F. Guerra 
\cite{bounds}, which relate the true quenched free energy to the Parisi solution with 
replica symmetry breaking \cite{MPV}.

Consider the system at temperature $\beta^{-1}$ and magnetic field $h$, and
recall that the Almeida-Thouless critical line is defined by the condition
\begin{equation}
\beta^2\int d\mu(z)\frac1{\cosh^{4}(z\beta \sqrt{\bar q}+\beta h)}=1,
\end{equation}
where $d\mu(z)$ is a  unit centered Gaussian measure and 
the Sherrington-Kirkpatrick order parameter $\bar q(\beta,h)$ is the
unique \cite{guerra1} solution of
\begin{equation}
\bar q=\int d\mu(z)\tanh^{2}(z\beta \sqrt{\bar q}+\beta h).
\end{equation}
The Parisi solution \cite{MPV} is defined as 
\begin{equation}
\bar\alpha_P(\beta,h)=\inf_{x\in \mathcal{X}} \bar\alpha(\beta,h;x),
\end{equation}
where $\mathcal{X}$ is the space of functional order parameters, i.e., of 
non decreasing functions $$x:q\in[0,1]\to x(q)\in
[0,1],$$ and $\bar\alpha(\beta,h;x)$ is defined as
\begin{equation}
\bar\alpha(\beta,h;x)=\ln2+f(0,h;x,\beta)-\frac{\beta^2}2\int_0^1 q\, x(q)dq.
\end{equation}
$f(q,y;x,\beta)$ is the solution of the antiparabolic equation
\begin{equation}
\partial_q f(q,y;x,\beta)+\frac12\left(\partial^2_y f(q,y;x,\beta)+x(q)
(\partial_y f(q,y;x,\beta))^2\right)=0
\end{equation}
with final condition
\begin{equation}
f(1,y;x,\beta)=\ln \cosh(\beta y).
\end{equation}
The equation for $f$ can be easily solved if $x(q)$ is piecewise constant.
For instance, if one takes
\begin{equation}
\left\{
\begin{array}{ll}
x(q)=0 & q\in[0,\bar q]\\
x(q)=1 & q\in(\bar q,1],
\end{array}
\right.
\end{equation}
one finds that $\bar\alpha(\beta,h,x)$ is the so called 
replica symmetric solution
\begin{equation}
\bar\alpha(\beta,h)=\ln2+\frac{\beta^2}4(1-\bar q)^2+
\int d\mu(z)\ln\cosh(z\beta\sqrt{\bar q}+\beta h).
\end{equation}

One expects the quenched free energy per site $F_N(\beta,h)$ to be related
to the Parisi solution by
\begin{equation}
-\lim_{N\to\infty}\beta\,F_N(\beta,h)=\bar\alpha_P(\beta,h),
\end{equation}
where $N$ is the size of the system. While the rigorous proof of this equality has not
yet been fully achieved, one can prove \cite{bounds} that
\begin{equation}
\label{bound}
-\beta\,F_N(\beta,h)\le\bar\alpha_P(\beta,h),
\end{equation}
for any value of $N,\beta,h$.

In the following, we employ the result (\ref{bound}) to prove that the 
thermodynamic limit of the quenched free energy is strictly greater 
than its replica symmetric approximation, below the Almeida-Thouless
line:
\begin{equation}
\label{risultato}
-\beta F(\beta,h)\equiv -\beta\lim_{N\to\infty}\,F_N(\beta,h)
<\bar\alpha(\beta,h),
\end{equation}
for 
\begin{equation}
\label{sotto}
\beta^2\int d\mu(z)\frac1{\cosh^{4}(z\beta \sqrt{\bar q}+\beta h)}>1.
\end{equation}
(The limit in (\ref{risultato}) exists, thanks to \cite{GTthermo}).
To this purpose, one simply needs to show that, if (\ref{sotto}) holds, there
exists a functional order parameter $\tilde x$ such that 
$\bar\alpha(\beta,h;\tilde x)<\bar\alpha(\beta,h)$.

For instance, we choose
\begin{equation}
\left\{
\begin{array}{ll}
\tilde x(q)=0 & q\in[0,\bar q]\\
\tilde x(q)=m & q\in(\bar q,q]\\
\tilde x(q)=1 & q\in(q,1],
\end{array}
\right.
\end{equation}
where $0\le m\le1$ and $\bar q\le q\le1$ and we denote with  
$\bar\alpha(\beta,h;m,q)$
the corresponding Parisi function $\bar \alpha(\beta,h;\tilde x)$.

Of course, since $\bar\alpha(\beta,h;1,q)=\bar\alpha(\beta,h)$,
it is sufficient to prove that
$$
\left.\partial_m\bar\alpha(\beta,h;m,q)\right|_{m=1}>0,
$$
for some $q$.
First of all,
$\bar\alpha(\beta,h;m,q)$ is easily found to be
\begin{eqnarray}
&&\bar\alpha(\beta,h;m,q)=\ln2+\frac{\beta^2}2(1-q)
-\frac{\beta^2}{4}(1-q^2+m(q^2-\bar q^2))+\\
&&+\frac1m\int d\mu(z')\ln\int d\mu(z)\cosh^m(\beta h+
\beta z\sqrt{q-\bar q}+\beta z'\sqrt{\bar q}).
\end{eqnarray}
Next, we compute the derivative with respect to $m$, keeping $q$ fixed, and 
we find
\begin{eqnarray}
&&\left.\partial_m\bar\alpha(\beta,h;m,q)\right|_{m=1}\equiv K(\beta,h;q)
\equiv\\\nonumber
&&\equiv-\frac{\beta^2}4(q^2-\bar q^2)-
\int d\mu(z')\ln\int d\mu(z)\cosh(\beta h+
\beta z\sqrt{q-\bar q}+\beta z'\sqrt{\bar q})\\\nonumber
&&+\int d\mu(z')\frac{\int d\mu(z)\cosh(\beta h+
\beta z\sqrt{q-\bar q}+\beta z'\sqrt{\bar q})\ln\cosh(\beta h+
\beta z\sqrt{q-\bar q}+\beta z'\sqrt{\bar q})}{\int d\mu(z)\cosh(\beta h+
\beta z\sqrt{q-\bar q}+\beta z'\sqrt{\bar q})}.
\end{eqnarray}
It is clear that, for $q\downarrow\bar q$, the integration over $z$ disappears, 
and $$K(\beta,h;\bar q)=0.$$
Therefore, in order to check the sign of $K(\beta,h;\bar q)$, we have to 
expand around $q=\bar q$.
By performing the first two derivatives with respect to $q$, one finds
$$\left.\partial_q K(\beta,h;q)\right|_{q=\bar q}=0$$
and
$$
\left.\partial^2_q K(\beta,h;q)\right|_{q=\bar q}
=-\frac{\beta^2}2\left(1-\beta^2
\int d\mu(z)\frac1{\cosh^4(z\beta \sqrt{\bar q}+\beta h)}\right).$$
This computation requires a simple integration by parts on a Gaussian 
variable. 
It is clear that, when condition (\ref{sotto}) holds, 
$\left.\partial^2_q K(\beta,h;q)\right|_{a=\bar q}>0$, so that
$$\left.\partial_m\bar\alpha(\beta,h;m,q)\right|_{m=1}>0,$$
at least for $q$ small.

This, together with Guerra's bound (\ref{bound}), completes the proof of 
the result (\ref{risultato}), i.e., of the instability of the replica 
symmetric solution.

\vspace{.5cm}
{\bf Acknowledgments}

The author is grateful to Francesco Guerra for many useful conversations.

This work was supported in part by MIUR 
(Italian Minister of Instruction, University and Research), 
and by INFN (Italian National Institute for Nuclear Physics).

\end{document}